\documentclass[aps,prl,reprint,amsmath,amssymb,floatfix,longbibliography,superscriptaddress]{revtex4-1}

\usepackage{graphicx}
\usepackage{amsmath,amssymb,bbold,bm,color}
\usepackage{siunitx}
\usepackage{float}
\usepackage{bbm}
\usepackage{epstopdf}
\usepackage{hyperref}
\usepackage[dvipsnames]{xcolor}
\usepackage[normalem]{ulem}

\usepackage[resetlabels,labeled]{multibib}

\newcites{New}{The other list}

\newcommand{\bk}{{\bm k}}

\newcommand{\bq}{{\bm q}}

\newcommand{\br}{{\bm r}}
\newcommand{\bv}{{\bm v}}

\newcommand\ii{\mathrm{i}}

\newcommand{\cT}{{\cal T}}

\newcommand{\bee}{\begin{equation}}
\newcommand{\ee}{\end{equation}}

\hypersetup{colorlinks=true,linkcolor=blue,citecolor=blue,urlcolor=blue}

\begin{document}

\title{Probing time reversal symmetry breaking topological superconductivity in twisted double layer copper oxides with polar Kerr effect}

\author{Oguzhan Can}
\affiliation{Department of Physics and Astronomy \& Stewart Blusson Quantum Matter Institute, University of British Columbia, Vancouver BC, Canada V6T 1Z4}

\author{Xiao-Xiao Zhang}
\affiliation{Department of Physics and Astronomy \& Stewart Blusson Quantum Matter Institute, University of British Columbia, Vancouver BC, Canada V6T 1Z4}
\affiliation{RIKEN Center for Emergent Matter Science (CEMS), Wako, Saitama 351-0198, Japan}

\author{Catherine Kallin}
\affiliation{Department of Physics and Astronomy, McMaster University, Hamilton, ON, Canada L8S 4M1}

\author{Marcel Franz}
\affiliation{Department of Physics and Astronomy \& Stewart Blusson Quantum Matter Institute, University of British Columbia, Vancouver BC, Canada V6T 1Z4}

\date{\today}

%
\begin{abstract}
  Recent theoretical work predicted emergence of chiral topological superconducting phase with spontaneously broken time reversal symmetry in a twisted bilayer composed of two high-$T_c$ cuprate monolayers, such as Bi$_2$Sr$_2$CaCu$_2$O$_{8+\delta}$. Here we identify large intrinsic Hall response that can be probed through the polar Kerr effect measurement as a convenient signature of the $\cT$-broken phase. Our modelling predicts the Kerr angle $\theta_K$ to be in the range of 10-100 $\mu$rad, which is a factor of  $10^3-10^4$ times larger than what is expected for the leading chiral supercondutor candidate Sr$_2$RuO$_4$. In addition we show that the optical Hall conductivity $\sigma_H(\omega)$ can be used to distinguish between the topological $d_{x^2-y^2}\pm id_{xy}$ phase and the $d_{x^2-y^2}\pm is$ phase which is also expected to be present in the phase diagram but is topologically trivial.
\end{abstract}

\date{\today}
\maketitle

{\em Introduction. --} While topological insulators and semimetals appear to be abundant in nature \cite{Lantagne2019} materials exhibiting topological superconductivity have been notoriously hard to find \cite{Kallin2016}.  The leading candidate for the spin triplet $p_x+ip_y$ chiral superconductor, $\text{Sr}_2\text{RuO}_4$, which showed time reversal symmetry breaking (TRSB) in muon spin resonance \cite{Luke1998} and polar Kerr effect measurements \cite{Xia2006}, has been recently shown to be inconsistent with triplet pairing by Knight shift experiment \cite{Pustogow2019}. Other suspected topological superconductors (TSCs) include FeTe$_x$Se$_{1-x}$ \cite{Zhang2018} and copper-doped Bi$_2$Se$_3$ \cite{Ando2011} but at present the balance of evidence remains inconclusive. Perhaps the most convincing case for TSC can be made in artificially engineered  proximitized semiconductor quantum wires where strong evidence for Majorana end modes has been reported by multiple groups and in a variety of experimental configurations \cite{Mourik2012,Das2012,Rokhinson2012,Finck2013,Deng2016}.

In this Letter, we consider a platform recently proposed to artificially engineer a chiral topological superconductor based on a pair of monolayer thick cuprate superconductors assembled with a twist \cite{Can2020,Volkov2020}. For a range of twist angles close to $45^{\rm o}$ such a bilayer is predicted to form a TRSB chiral phase that can be thought of as a  $d_{x^2-y^2}\pm id_{xy}$ superconductor (which we abbreviate as $d\pm id'$ henceforth).  We address here the pressing question of how such a topological $\cT$-broken phase can be reliably identified in experiment. Specifically, we show that owing to its multiband nature and broken symmetries such a system exhibits strong anomalous Hall conductivity which can be probed optically through the polar Kerr effect (PKE) measurement.

Finite magneto-optic Kerr effect is a manifestation of time reversal symmetry breaking \cite{Kapitulnik2009} in a material where linearly polarized incident light is reflected with elliptical polarization whose major axis is rotated with respect to the incident polarization axis. Kerr angle $\theta_K$ is directly related to the anomalous Hall conductivity $\sigma_H(\omega)$ defined below. TRSB is a necessary but not sufficient condition for finite Kerr rotation. In addition, any mirror reflection symmetry perpendicular to the incident wave vector must also be broken for $\sigma_H(\omega)$ to be nonzero \cite{Wang2014}. In this work, we calculate the anomalous Hall conductivity of two stacked CuO$_2$ monolayers twisted with respect to one another by an angle close to 45 degrees. Such a configuration develops a complex phase between the $d$-wave order parameters in the two layers, spontaneously breaking the time reversal symmetry as well as all mirror symmetries. The resulting system can be in a $d+id'$ or $d+is$ phase, depending on parameters such as doping and interlayer coupling strength \cite{Can2020}. Both phases break $\cT$ and are  fully gapped but only the former exhibits topologically protected chiral Majorana edge modes. In this phase we find a large intrinsic contribution  to $\sigma_H(\omega)$ arising from interband transitions that involve chiral Cooper pairs -- a mechanism originally identified in Ref.\ \cite{Kallin2012}.


{\em PKE and optical Hall conductivity. --}
In the presence of time reversal symmetry breaking, the polarization axis of a linearly polarized incident light will be rotated by the Kerr angle $\theta_K$ given by \cite{Argyres1955}
\begin{equation}\label{eq:kerrangle}
  \theta_K(\omega)=\frac{4\pi}{2d\omega}\text{Im}\left(\frac{\sigma_H(\omega)}{n(n^2-1)}\right)
\end{equation}
where $n$ is the complex index of refraction, $d$ denotes the separation of monolayer pairs and $\sigma_H(\omega)$ is the antisymmetric part of the optical Hall conductivity per monolayer
\begin{equation}\label{hallconductance}
    \sigma_H(\omega)=\frac{1}{2}\lim_{\textbf{q}\rightarrow 0}\left[\sigma_{xy}(\textbf{q},\omega)-\sigma_{yx}(\textbf{q},\omega)\right].
\end{equation}
Eq.\ \eqref{eq:kerrangle} is valid when the sample thickness $h\gg \lambda$, the wavelength of the incident light. As such it would apply to a stack composed of many bilayers.  In the opposite limit  $h\ll\lambda$, relevant to a single bilayer, the Kerr angle is given by a different expression \cite{Tse2011} which in addition involves the longitudinal optical conductivity $\sigma_{xx}(\omega)$ and is specified in Eq.\ \eqref{eq:thinfilm_kerrangle} below. Crucially, in both limits $\theta_K$ can be non-zero only when the Hall conductivity is finite. 
The latter can be computed from the Kubo formula as $\sigma_{xy}(\textbf{q},\omega)=\frac{1}{\omega} (\pi_{xy}(\textbf{q},\omega) - \pi_{xy}(\textbf{q},0))$.
Here $\omega$ is the frequency of the incident light and
\begin{equation}\label{eq:jxjycorrelator}
    \pi_{xy}(\textbf{q},\omega)=\int_0^\infty dt e^{i\omega t}\langle[\hat{J}_x^\dag(\textbf{q},t),\hat{J}_y(\textbf{q},0)]\rangle
\end{equation} 
is the current-current correlator. 
The total current operator $\hat{J}_i=e\sum_k \text{tr}\Psi_\textbf{k}^\dag \hat{\textbf{v}}_i \Psi_\textbf{k}$ uses $\hat{\textbf{v}}_i = (\mathbb{1} \otimes \sigma_z)\partial_{k_i}h^0_\textbf{k}$ where  $h^0_\textbf{k}= h_\textbf{k} - h^\Delta_\textbf{k}$ is the normal part of the system Bloch Hamiltonian $h_\textbf{k}$ expressed in the orbital basis and $\sigma_z$ acts in the particle-hole (Nambu) space.
Under time reversal the current correlator transforms as $\pi_{xy} \rightarrow \pi_{yx}$ while under mirror reflections along $x$ or $y$ axes $\pi_{xy} \rightarrow -\pi_{xy}$. 
Therefore, in order to obtain a nonzero Kerr effect, both time reversal and all mirror symmetries must be broken \cite{Wang2014}.

Equations (\ref{hallconductance},\ref{eq:jxjycorrelator})  can be rewritten in terms of the eigenspectrum of $h_\textbf{k}$ as
\begin{align}\label{eq:sigma_H_comp}
 \sigma_H(\omega)=&\frac{i e^2}{\omega} \sum_{\textbf{k},ab}  \frac{(\omega + i\epsilon)Q_{ab}}{(E^a_{\textbf{k}} - E^b_{\textbf{k}})^2 - (\omega + i\epsilon)^2 } n_F(E^a_{\textbf{k}})   
\end{align}
where $a,b$ are band indices and $Q_{ab} = 2i\,\text{Im}\{v^x_{ab} v^y_{ba}\}_\textbf{k}$ with the matrix element $v^j_{ab}=\langle a{\textbf{k}}\lvert \hat{\textbf{v}}_j\rvert b{\textbf{k}}\rangle$ computed from the Bloch eigenstates. This formulation is convenient for numerical calculations based on the lattice models of twisted bilayers discussed below. Details of the derivation are given in the Supplementary Material (SM).
\begin{figure}[t]
\includegraphics[width=1\columnwidth]{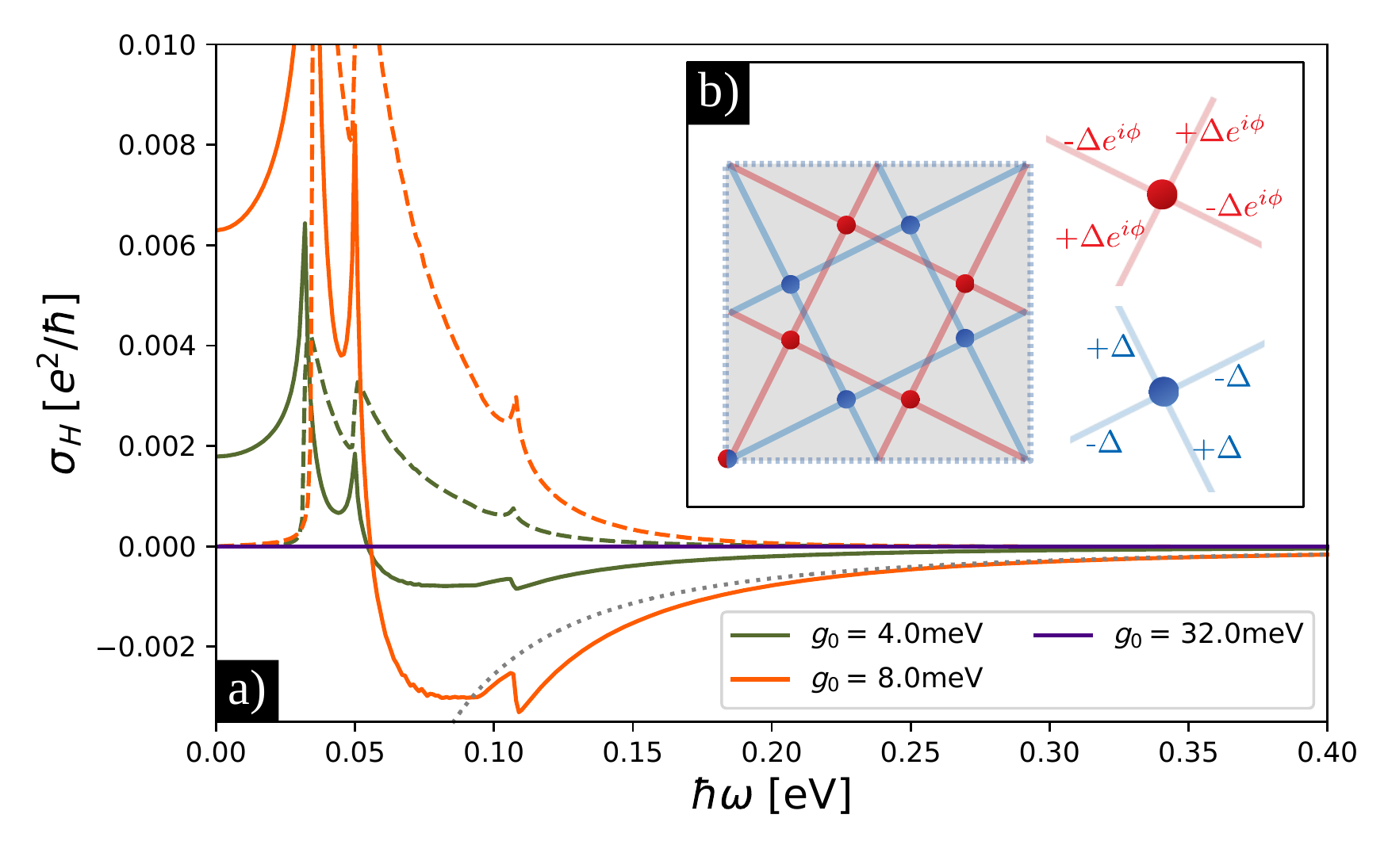}
\caption{a) Optical Hall conductivity for the 10-band lattice model calculated from Eq.\ \eqref{eq:sigma_H_comp} for various values of the interlayer coupling strength $g_0$. Solid and dashed lines show the real and the imaginary parts of  $\sigma_H(\omega)$, respectively. The tight binding model parameters are chosen to be $t=0.153$eV, $t'= -0.45t$, $\mu=-1.3t$. The SC order parameter is calculated self consistently with $V_{ij}$ chosen such that the maximum gap is 40meV. Grey dotted curve shows the $\omega^{-2}$ scaling which is the  expected behavior of  ${\rm Re}\sigma_H(\omega)$ in the high frequency limit.
b) The 10-band model unit cell with two stacked layers depicted in blue and red. 
The signs characteristic of $d$-wave  order parameter along with the relative complex phases are indicated in the right panel. Note that the complex phases in  the order parameters break the mirror reflection symmetries along the $x$, $y$ and $x\pm y$ directions.}
\label{fig:conductance-10band-model}
\end{figure}

{\em 10-band model of coupled twisted copper oxide layers. --}
Following Ref.\ \cite{Can2020}  we consider two stacked cuprate monolayers twisted with respect to each other by a commensurate twist angle $\theta_{m,n}=2 \arctan{(m/n)}$. Each CuO$_2$ plane is modeled by a minimal  Hubbard model with nearest neighbor attractive potential, known to produce a $d_{x^2-y^2}$ superconductor within the standard Bogoliubov-de Gennes (BdG) theory.   The Hamiltonian has the form 
\begin{eqnarray}
  H&=& - \sum_{ ij, \sigma a} t_{ij}c^\dag_{i \sigma a} c_{j \sigma a} 
  - \mu \sum_{i \sigma a} n_{i \sigma a} \nonumber \\
  &+&\sum_{ ij,a}V_{ij}n_{ia}n_{ja}
- \sum_{i j \sigma} g_{ij} c^\dag_{i \sigma 1} c_{j \sigma 2},
    \label{hm_latt}
\end{eqnarray} 
where $t_{ij}$ denote the nearest and next nearest neighbor intralayer tunneling amplitudes between sites $i$ and $j$. Chemical potential $\mu$ controls the doping, $V_{ij}$ are the attractive interactions. The interlayer coupling strengths $g_{ij}$ are assumed to decay exponentially with distance $r_{ij} = \sqrt{d^2 + \rho_{ij}^2}$ as
\begin{equation}\label{eq:10band_interlayercouplingform}
	g_{ij} = g_0 e^{-(r_{ij}-d)/\zeta},
\end{equation}
where $\rho_{ij}$ is the in-plane separation between sites $i$ and $j$, $d$ is the distance between the monolayers and $\zeta$ denotes the characteristic lengthscale for interlayer tunneling.

For simplicity and concretness we will focus here on a specific commensurate twist angle  $\theta_{1,2}=2\arctan{(1/2)} \simeq 53.13^{\rm o}$ which leads to the smallest nontrivial moir\'{e} unit cell with 10 sites, as sketched in Fig.\ \ref{fig:conductance-10band-model} and is sufficiently close to  $45^{\rm o}$ to illustrate all the interesting physics. We will refer to this lattice model as the 10-band model in the following. In SM we show Kerr angle results for other commensurate twist angles that are computationally acessible.  

\begin{figure}[t]
\includegraphics[width=1\columnwidth]{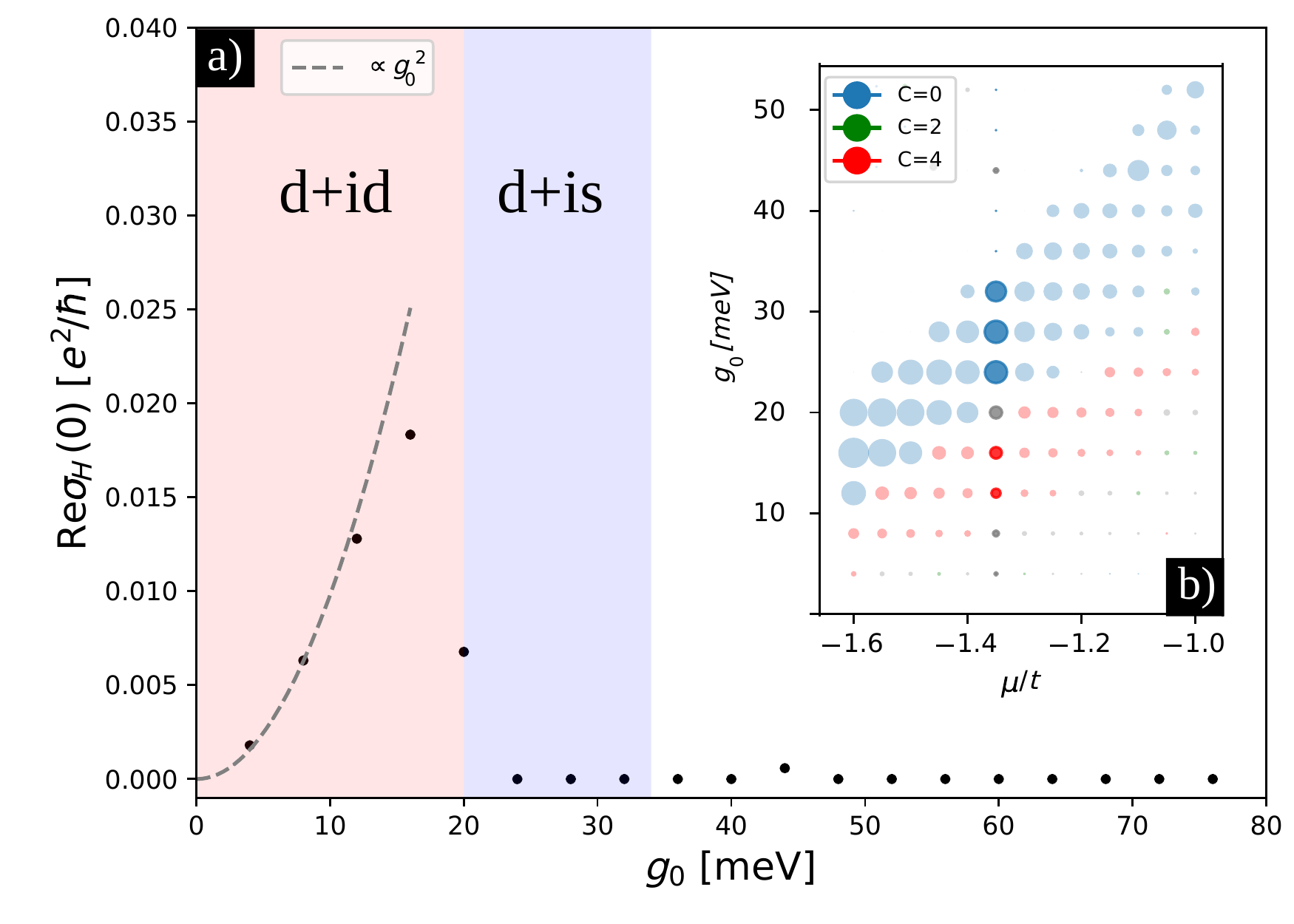}
\caption{a) Zero frequency limit of the real part of the optical Hall conductivity $\sigma_H(\omega)$ for the 10-band lattice model for the same parameters as Fig.\  \ref{fig:conductance-10band-model}. The data shown in the main panel corresponds to the highlighted slice in the inset (b) which displays the phase diagram of this configuration as a function of $g_0$ and $\mu/t$. The color corresponds to the Chern number $C$ while the radius of the dots is proportional to the minimum excitation gap.}
\label{fig:tuning_g}
\end{figure}

The phase diagram of the 10-band model, obtained through the standard mean-field decoupling of the interaction term in the pairing channel and then solving the resulting gap equations \cite{Can2020}, is displayed in Fig.\ \ref{fig:tuning_g}(b).
At weak interlayer coupling $g_0$ we find that $d$-wave superconducting order parameters in twisted monolayers acquire a complex relative phase, spontaneously breaking the time reversal symmetry and forming  a $d+id'$ TSC. This phase is characterized by  Chern number $C=4$ which can be understood as two layers of $d+id'$ TSC, each contributing $C=2$.   As $g_0$ gets stronger the system transitions into the $\cT$-broken but topologically trivial  $d+is$ state with $C=0$.  For even larger $g_0$ the complex phase between layers disappears resulting in a gapless phase that can be thought of as two independent $d_{x^2-y^2}$ superconductors in the band basis. We note that for twist angles closer to $45^{\rm o}$ the gapless phase is replaced by another TSC phase characterized by $C=2$ \cite{Can2020} -- this can be regarded as a single-layer $d+id'$ superconductor in the band basis.

Fig.\ \ref{fig:conductance-10band-model} shows our results for the Hall conductivity $\sigma_H(\omega)$ in the 10-band model computed using Eq.\ \eqref{eq:sigma_H_comp}. The parameters are chosen to capture Bi$_2$Sr$_2$CaCu$_2$O$_{8+\delta}$  near optimal doping where we expect the mean-field BCS theory to provide an accurate description of the superconducting state.
Because the strength of the interlayer coupling  $g_0$ is difficult to accurately estimate for the twisted configuration we display $\sigma_H(\omega)$ for several representative values spanning its likely range.
When $g_0<20$meV the ground state has Chern number $C=4$. The Hall conductivity is non-zero with an amplitude that is {\em three to four orders} of magnitude larger than the prediction for $\text{Sr}_2\text{RuO}_4$ given in Ref.\ \cite{Kallin2012}. We show below that this leads to giant Kerr angle that should be straightforward to detect experimentally.
By contrast in the $C=0$ phase, obtained for $g_0=32$meV,  $\sigma_H(\omega)$ is much smaller ($\approx 10^{-7} e^2/\hbar $) and will lead to very weak Kerr signal.

{\em Effective 2-band model. --}
In order to gain some insight into the Hall response of twisted bilayers we now consider an effective 2-band model designed to capture the essential features of the system. The 2-band BdG Hamiltonian $h_\bk^{\rm eff}$ and its derivation are given in SM. 
The Hall conductivity Eq.~\eqref{hallconductance} for the 2-band model is found to be
\begin{equation}\label{eq:sigmaH_2band}
\begin{split}
  &\sigma_H(\omega)=2e^2 \sum_\bk \mathrm{Im}[\Delta_1^*\Delta_2]\,
[\partial_\bk(\xi_2-\xi_1)\times \partial_\bk |g|^2]_z \times \\
    & \frac{1}{E_1E_2} \left[ \frac{1-n_F(E_1)-n_F(E_2)}{E_+(E_+^2-(\omega+i\epsilon)^2)} - \frac{n_F(E_1)-n_F(E_2)}{E_-(E_-^2-(\omega+i\epsilon)^2)} \right],
\end{split}
\end{equation}
where $\xi_{1(2)}$ and $\Delta_{1(2)}$ are the effective dispersions and gap functions defined in Eqs.\ (\ref{eq:disp_twoband},\ref{eq:pairing_twoband}) of SM, 
$E_{1,2}$ denote  the positive  energy eigenvalues of $h_\bk^{\rm eff}$ and $E_\pm=E_1\pm E_2$. SM gives the derivation of Eq.\ \eqref{eq:sigmaH_2band} and also shows comparison with $\sigma_H(\omega)$ obtained directly from the 10-band model.

Eq.~\eqref{eq:sigmaH_2band} is useful because it clarifies conditions under which a system can exhibit non-vanishing anomalous Hall effect. Specifically, three physical ingredients are required, in addition to broken $\cT$ and mirror symmetries: (i) the two normal-state  dispersions must be different $(\xi_1\neq\xi_2)$, (ii) a non-vanishing SC phase difference must exist between the layers such that  $\mathrm{Im}[\Delta_1^*\Delta_2]$ is non-zero, and (iii) the interlayer coupling $g$ must be $\bk$-dependent. In the two-band model defined by Hamiltonian \eqref{eq:bdgham_twoband} the first two conditions are satisfied for a $d+id'$ state and when second-neighbor in-plane tunneling $\tilde{t'}$ is non-zero. The third condition will be met if we allow interlayer hopping between sites that are not directly above one another. 
For instance we may assume the two square lattices to be offset by $(b/2,b/2)$ and allow nearest neighbor hopping; this gives $g(\bk)=4g_0\cos{(k_xb/2)}\cos{(k_yb/2)}$. Importantly, we expect the above conditions to be met in a physical sample of a twisted Bi$_2$Sr$_2$CaCu$_2$O$_{8+\delta}$ bilayer.

To produce finite and significant Hall signal, the integrand in Eq.~\eqref{eq:sigmaH_2band} must be even under the reflection $x\leftrightarrow y$. This is most easily ensured when all factors are even: the $\cT$-breaking $\mathrm{Im}[\Delta_1^*\Delta_2]$, the vertex, and the energy factors are all even in the $d+id'$ case. On the other hand  $\mathrm{Im}[\Delta_1^*\Delta_2]$ becomes odd in the $d+is$ case which renders the signal vanishing. This exemplifies a situation that breaks both mirror and time-reversal symmetries but still exhibits no Kerr signal. This distinguishing feature between the two $\cT$-breaking cases, trivial $d+is$ and topological $d+id'$, is indeed confirmed in the 10-band calculation in Fig.~\ref{fig:tuning_g}, where the former exhibits negligibly small signal compared to the latter.

At zero temperature, only the first term on the second line of Eq.~\eqref{eq:sigmaH_2band} contributes. This term describes a process of breaking a  Cooper pair into two quasiparticles on different branches of the spectrum and indicates the onset of the absorptive part $\mathrm{Im}[\sigma_H(\omega)]$ at $\omega^*=\min_\bk[E_1(\bk)+E_2(\bk)]$. Note that  $\omega^*$ is roughly set by the energy scale of the interlayer coupling $g$ when the pairing is small compared to the chemical potential and is in general much larger than the quasiparticle gap minimum. We expect this physical picture and the form of $\omega^*$  to hold approximately in the 10-band model with $E_{1,2}$ being the two lowest energy levels. This is indeed confirmed in our numerical calculations shown in Fig.~\ref{fig:conductance-10band-model} and SM Fig.\ \ref{fig:conductance-comparison}. 

Lastly, Eq.~\eqref{eq:sigmaH_2band} shows that the Hall signal amplitude exhibits  quadratic dependence on $g$ and $\Delta$ to leading order. This is numerically confirmed to hold in the 10-band model as indicated in Fig.~\ref{fig:tuning_g} for the $g$-dependence. This observation suggests a plausible explanation for why  Fig.\ \ref{fig:conductance-10band-model} shows signal four orders of magnitude larger than that predicted for Sr$_2$RuO$_4$ \cite{Kallin2016}: in Bi$_2$Sr$_2$CaCu$_2$O$_{8+\delta}$ the gap $\Delta$ is about two orders of magnitude larger while $g$ is roughly the same in the two materials. 

{\em Kerr angle estimate. --} We use Eq.\ \eqref{eq:kerrangle} and its  
thin sample ($\lambda \gg h$) counterpart \cite{Tse2011} given by
\begin{equation}\label{eq:thinfilm_kerrangle}
    \theta_K=\text{Re}\arctan\left(\frac{-\sigma_{xy}}{\sigma_{xx}+4\pi(\sigma_{xx}^2+\sigma_{xy}^2)}\right)
\end{equation}
to estimate the expected Kerr angle.  
Here the optical conductivities are made dimensionless by attaching the fine structure constant $\alpha$, i.e., setting $e^2/\hbar = \alpha$ in the natural units.
The C$_4$ rotation symmetry of the twisted bilayer implies $\sigma_{xy}=-\sigma_{yx}$ which, in combination with Eq.~\eqref{hallconductance}, gives $\sigma_{xy}=\sigma_H$.
 The diagonal part of the conductivity tensor and the complex index of refraction required in Eq.\ \eqref{eq:kerrangle} are dominated by various relaxation mechanisms whose microscopic origin is poorly understood in the cuprates.  To estimate $\sigma_{xx}$ 
we thus rely on an empirical power-law formula 
\begin{equation}\label{eq:timusk}
\sigma_{xx} \simeq 2d C(-i\omega)^{\gamma-2}, \ \ \ \gamma=1.447,
\end{equation}
that was shown in Ref.\ \cite{Hwang2007} to accurately describe the ab-plane reflectance data on Bi$_2$Sr$_2$CaCu$_2$O$_{8+\delta}$ for frequencies above $\omega_c\simeq 0.12$eV at all temperatures. System thickness $h=2d$ is inserted to convert between the bulk and thin film conductivity.

\begin{figure}[t]
\includegraphics[width=1\columnwidth]{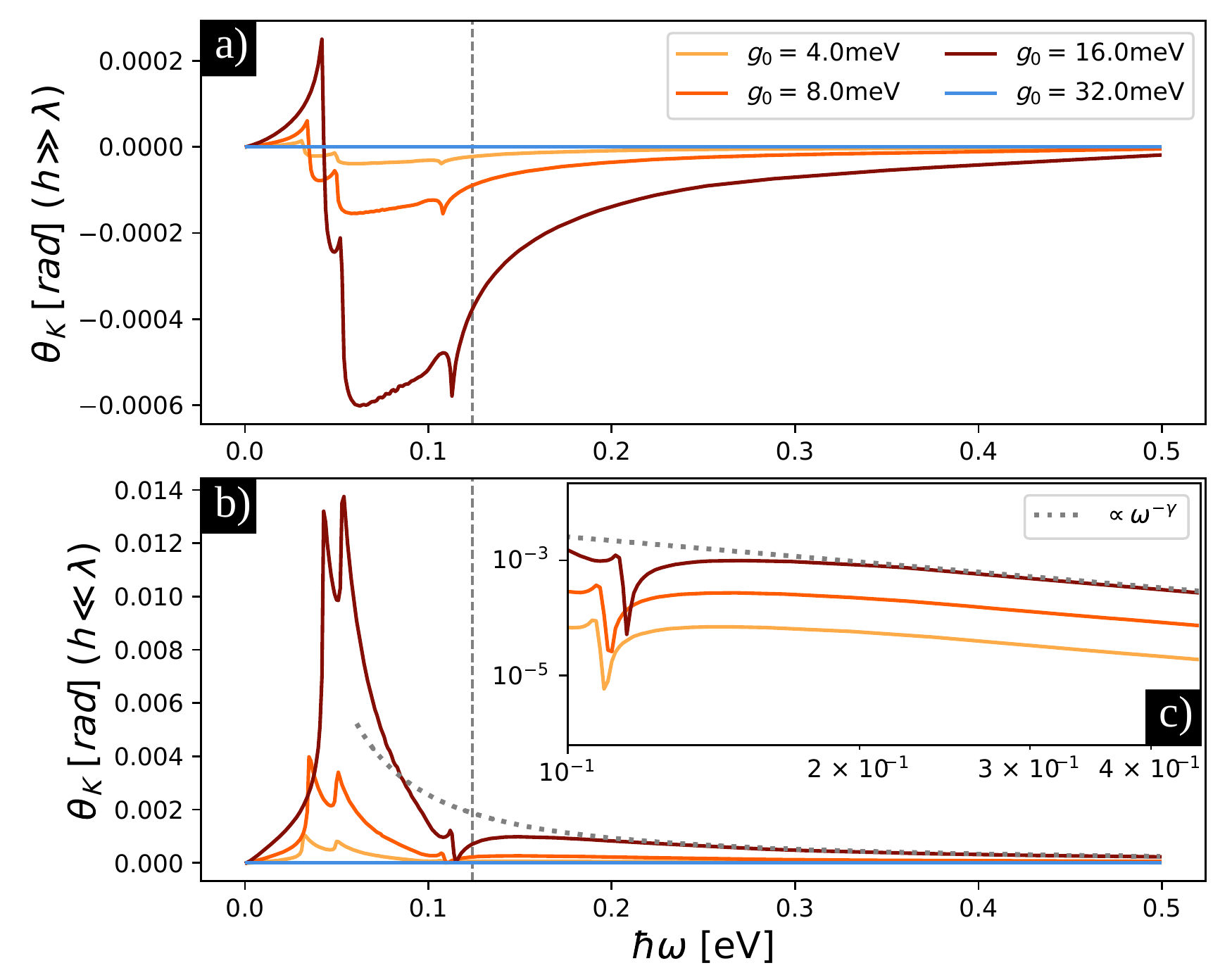}
\caption{Kerr angle $\theta_K$ as a function of photon energy $\hbar\omega$ for various values of interlayer coupling $g_0$. a) Thick sample limit Eq.\ \eqref{eq:kerrangle}. b) Thin sample limit Eq.\ \eqref{eq:thinfilm_kerrangle}. c) Semi-log plot with the expected  $\omega^{-\gamma}$ scaling at high frequencies shown by the dotted grey line. Vertical dashed lines indicate the frequency $\omega_c$ above which $\sigma_{xx}$ shows the power-law form $\propto (-i\omega)^{\gamma-2}$}
\label{fig:kerrangle_main}
\end{figure}

Fig.\ \ref{fig:kerrangle_main} shows our results for PKE with further details provided in SM. For simplicity we assume the power law Eq.\ \eqref{eq:timusk} to be valid at all frequencies and therefore expect our predictions for $\theta_K$ to be less accurate for $\omega\lesssim \omega_c$. We note, however, that PKE experiments are typically performed at frequencies above $\sim 0.5$ eV  \cite{Xia2006,Kapitulnik2009}, well within the range of applicability of Eq.\ \eqref{eq:timusk}. 

In this experimentally relevant regime it is possible to obtain a simple approximate expression for $\theta_K$ in the thin-sample limit that can be used as guidance in experimental studies. As shown in SM for $\omega\gtrsim 0.12$eV one may approximate Eq.\ \eqref{eq:thinfilm_kerrangle} as
\begin{equation}\label{scaling}
    \theta_K \simeq \Lambda g_0^2\omega^{-\gamma}, 
\end{equation} 
where $\theta_K$ is in radians and $\Lambda = 0.3623 [eV]^{\gamma-2}$ is a constant we extract by fitting the curves in Fig.\ \ref{fig:kerrangle_main}, which show excellent agreement with the scaling form \eqref{scaling}. For the typical photon frequency $\hbar\omega=0.5$eV Eq.\ \eqref{scaling} predicts $\theta_K\simeq 10-100$ $\mu$rad if we assume $g_0=4-16$ meV.  

{\em Summary and conclusions. --} Our modelling predicts large {\em intrinsic} contribution to the anomalous Hall conductivity of twisted Bi$_2$Sr$_2$CaCu$_2$O$_{8+\delta}$ bilayer in the spontaneously $\cT$-broken topological phase, which is expected to occur for a range of twist angles close to $45^{\rm o}$. There could be other `extrinsic' contributions to $\sigma_H(\omega)$ from various scattering mechanisms discussed in the literature \cite{Yip1992,Roy2008,Goryo2008,Lutchyn2009} which we have not considered here. Typically, such contributions arise from higher-order diagrams in the expansion of the current-current correlator Eq.\ \eqref{eq:jxjycorrelator} and are thus subdominant with respect to the intrinsic component identified here. To the extent that various contributions are additive our estimate for the Kerr angle should therefore be viewed as a lower bound. Its large magnitude, measured in $\mu$rad rather than nrad, typical of Sr$_2$RuO$_4$ and other chiral SC candidates such as UPt$_3$ \cite{Schemm2014,Kallin2017}, gives hope that spontaneous $\cT$ breaking in twisted high-$T_c$ cuprate bilayers can be reliably and unambiguously detected with existing laboratory instruments.

{\em Acknowledgements. --}
We thank D.A.\ Bonn, A. Damascelli, E. Ostroumov, Yunhuan Xiao and Ziliang Ye for useful discussions and correspondence. This work was supported by NSERC, the Max Planck-UBC-UTokyo Centre for Quantum Materials and the Canada First Research Excellence Fund, Quantum Materials and Future Technologies Program.  OC is supported by International Doctoral Fellowship from UBC.

\bibliography{PKE}

\newpage

\pagebreak

\setcounter{equation}{0}
\setcounter{figure}{0}
\setcounter{table}{0}
\setcounter{page}{1}
\makeatletter
\renewcommand{\theequation}{S\arabic{equation}}
\renewcommand{\thefigure}{S\arabic{figure}}

\section{APPENDIX}

\subsection{Details on the 2-band model}
In this section we give a derivation of the 2-band model discussed in the main text.  We assume the regime of strong hole doping such that the Fermi surface is small and confined to the center of the Brillouin zone $\Gamma$. While this may be relevant for strongly overdoped Bi$_2$Sr$_2$CaCu$_2$O$_{8+\delta}$ the primary goal of this derivation is to obtain a simple, analytically tractable model, that will provide insights into the physics of the Hall conductivity in the general class of twisted bilayer $d$-wave superconductors.

To construct the 2-band model we employ the following strategy. We  begin from two initially decoupled CuO$_2$ layers described as BCS $d$-wave  superconductors. We expand the electron dispersions and gap functions to leading order in small momentum $k$ around the $\Gamma$ point and perform a rotation by angle $\pm \theta/2$ in the two layers. Next we regularize these expressions on two perfectly aligned (i.e.\ unrotated) square lattices with the lattice constant $b$ given by the moir\'{e} unit cell and write down the corresponding tight-binding model. Finally we couple the two layers by a generic single-electron tunneling $g(\bk)$ consistent with the system symmetries. The final 2-band model is thus defined on a pair of aligned square lattices with the twist encoded in the effective dispersion relations and gap functions.  
\begin{figure}[b]
    \includegraphics[width=\columnwidth]{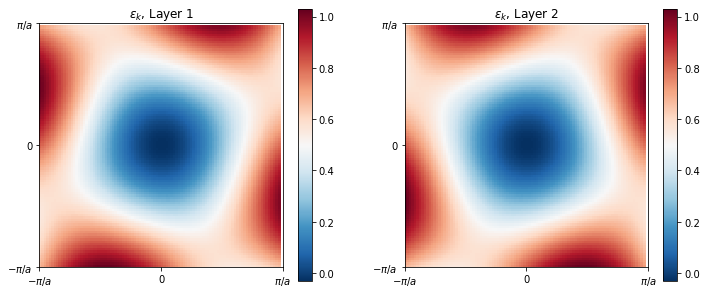}
        \includegraphics[width=\columnwidth]{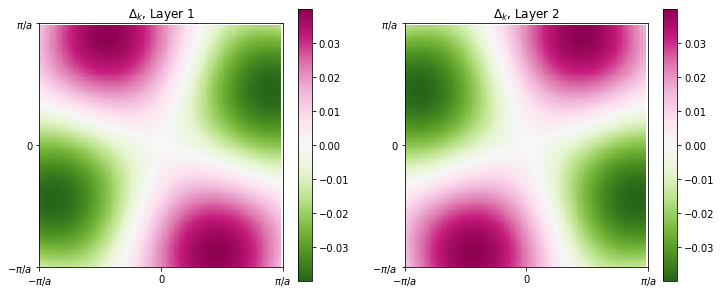}
    \caption{Normal parts of  energy dispersions (top) and the order parameter (bottom) for decoupled layers that are naively rotated  by $\pm \theta/2$ with $\theta=2\arctan(1/2)$. We observe that the the patterns are no longer periodic when plotted  in the original Brillouin zone.}
    \label{fig:singlelayerdispersions1}
  \end{figure}

{\em $d+id$ case. --}
Starting with two decoupled CuO$_2$ planes we expand the dispersion of each monolayer near the BZ center, keeping terms up to $\mathcal{O}(k^4)$
\begin{align}
 \epsilon_k &= -2t(\cos k_xa + \cos k_ya) \\
            & \simeq -2t\left[2 - \frac{1}{2}k^2a^2 +  \frac{1}{8}k^4a^4 - \frac{1}{16}(2k_xk_y)^2a^4\right]. \nonumber
\end{align}
The fourth order terms are required to capture the 4-fold anisotropy of the dispersion which, as we shall see, turns out to be important for the Hall response. The pair functions are expanded to second order
\begin{align}
 \Delta_k = \frac{\Delta_0}{2}(\cos k_xa - \cos k_ya) \simeq -\frac{\Delta_0}{4}(k_x^2-k_y^2)a^2.
\end{align}
\begin{figure}[t]
    \includegraphics[width=\columnwidth]{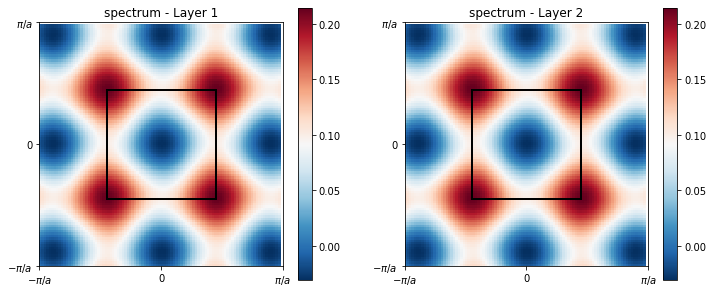}
        \includegraphics[width=\columnwidth]{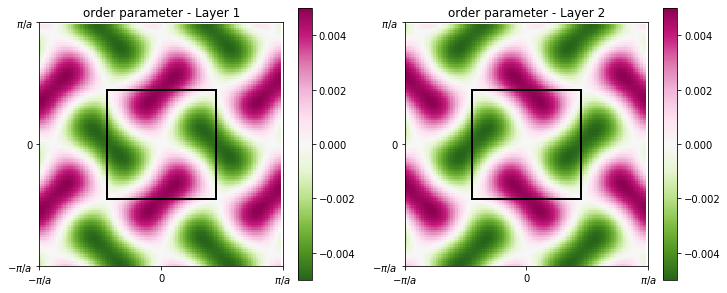}    \includegraphics[width=\columnwidth]{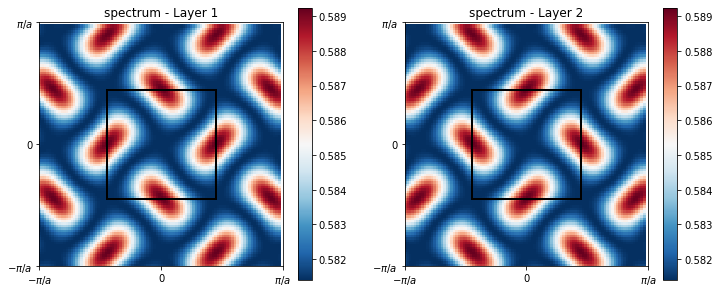}
        \caption{Normal state dispersion (top) and order parameter (middle) of the effective 2-band model constructed for $\theta=2\arctan(1/2)$. The black square encloses the moir\'{e} unit cell with length $2\pi/b$ and we observe that the patterns are now properly periodic. The last term in Eq.\ \eqref{s3}  is plotted separately (bottom row) as it is very small compared to the other terms. However, inclusion of this term is very important since without it we would have $\xi_1=\xi_2$ and the Hall conductivity would identically vanish. }
    \label{fig:singlelayerdispersions}
\end{figure}
Now we account for the twist by counter-rotating each layer by angle $\pm\theta/2$
as shown in Fig. \ref{fig:singlelayerdispersions1}.
Under a rotation by $\theta/2$ various terms transform as
$$
k_x^2-k_y^2 \rightarrow (k_x^2-k_y^2)\cos\theta + (2k_xk_y)\sin\theta ,
$$
$$
2k_xk_y \rightarrow (2k_xk_y) \cos\theta  - (k_x^2-k_y^2)\sin{\theta}.
$$

Next we construct a tight binding model whose low-energy expansion near the $\Gamma$ point gives the above rotated dispersion relation and gap functions.
This is achieved by replacing
$$
k_xa \rightarrow \frac{a}{b}\sin(k_x b ), \ \ \
(k_xa)^2 \rightarrow {\frac{2a^2}{ b^2}}[1-\cos{(k_x b)}],
$$
and similar for $k_y$. Here $b$ is the lattice constant of the moir\'{e} lattice that results from the twisted geometry (we assume commensurate twist angle $\theta$).   We thus obtain
\begin{eqnarray}
  \xi_k(\theta)=&-&\mu -4t\left(1-a^2/b^2\right) \hspace{120pt} \label{s3} \\
                &-&2t\biggl[ \frac{a^2}{b^2}(\cos k_xb + \cos k_y b)\nonumber \\
  &-& \frac{1}{4}\frac{a^4}{b^4}\left(\eta_{xy}\cos \theta + \eta_{x^2-y^2} \sin \theta\right)^2\biggr], \nonumber
\end{eqnarray}
and the pairing
\begin{equation}\label{s4}
\Delta_k(\theta) = \frac{\Delta_0}{2}\frac{a^2}{b^2}\left[ \eta_{x^2-y^2} \cos\theta - \eta_{xy} \sin \theta\right],
\end{equation}
where
\begin{align}
  \eta_{xy}=&\sin{k_x b} \sin{k_y b},  \\
  \eta_{x^2-y^2}=&\cos{k_x b} - \cos {k_y b}.
\end{align}
The resulting dispersions can be seen in Fig.\ \ref{fig:singlelayerdispersions}.

If we define $\tilde{\mu}=\mu + 4t\left(1-a^2/b^2\right)$, 
$\tilde{t}=ta^2/b^2$, 
$\tilde{t'}=\frac{t}{2}a^4/b^4$ 
and
$\tilde{\Delta}=\Delta_0(a^2/b^2)$ as the effective parameters, we obtain the 2-band model 
defined by the BdG Hamiltonian $\mathcal{H}=\sum_\bk \Psi_\bk^\dag h_\bk^{\rm eff} \Psi_\bk$ with
\begin{equation}\label{eq:bdgham_twoband}
  h_\textbf{k}^{\rm eff}= \begin{pmatrix} \xi_1(\bk) & \Delta_{1}(\bk) & g(\bk) & 0 \\
    \Delta_{1}(\bk)^\ast & -\xi_1(\bk) & 0 & -g(-\bk)^\ast \\
     g(\bk)^\ast & 0 & \xi_2(\bk) & \Delta_{2}(\bk) \\
    0 & -g(-\bk) & \Delta_{2}(\bk)^\ast & -\xi_2(\bk) \end{pmatrix}
\end{equation}
and the Nambu spinor $\Psi_\bk=(c_{\bk\uparrow
1},c^\dag_{-\bk\downarrow 1}, c_{\bk\uparrow 2},c^\dag_{-\bk\downarrow 2})^{T}$. 
The effective dispersions and gap functions are given by
\begin{eqnarray}
  \xi_{1(2)}(\bk)&=&-\tilde{\mu} -2\tilde{t}(\cos{k_x b} + \cos{k_y b}) \label{eq:disp_twoband} \\
  &+& \tilde{t'}\left(\eta_{xy}\cos \theta \pm \eta_{x^2-y^2} \sin \theta\right)^2 \nonumber \\
\Delta_{1(2)}(\bk) &=& \frac{\tilde{\Delta}}{2}\left(\eta_{x^2-y^2} \cos\theta \mp \eta_{xy} \sin \theta\right]e^{\pm i\phi/2} \label{eq:pairing_twoband}
\end{eqnarray}
The requisite $\cT$ breaking is encoded through the complex phase difference $e^{i\phi}$  between superconducting order parameters in two layers.

In relation to the 10-band model the lattice constant is $b=\sqrt{1^2+2^2}a=\sqrt{5}a$. The corresponding BZ will be smaller, with the dimension $2\pi/b$. The 10-band model can be recovered from the 2-band model  by folding the dispersions in Fig.\ \ref{fig:singlelayerdispersions} back into the smaller moir\'{e} BZ. If we are interested in features that are near the center of the BZ and small enough energies such that they do not backfold, the 2-band model with the above parameters should provide an accurate description.

{\em $d+is$ case. --}
For the $d+is$ case, the pairing form obtained from self-consistent calculation is
\begin{equation}
    \Delta_{1(2)}(k) = \pm \Delta^{d} \eta_{x^2-y^2}  - i \Delta^{s} \chi
 \end{equation}
in the local frame of each layer with $\Delta^{d}=\tilde{\Delta} e^{i\phi/2}\cos{\phi/2}$, $\Delta^{s}=\tilde{\Delta} e^{i\phi/2}\sin{\phi/2}$ and $\chi=\cos k_x b + \cos k_y b$. We can follow the transformation procedure described above to obtain
\begin{eqnarray}
    \Delta_{1(2)}(k) = &\pm& \Delta^{d} (\eta_{x^2-y^2} \cos{\theta} \mp \eta_{xy}\sin{\theta}) \\
    & -& i\Delta^{s} [\chi - \frac{1}{4}(\eta_{xy}\cos{\theta} \pm \eta_{x^2-y^2}\sin{\theta})^2]\nonumber.
\end{eqnarray}
%


\begin{figure}[t]
\includegraphics[width=1\columnwidth]{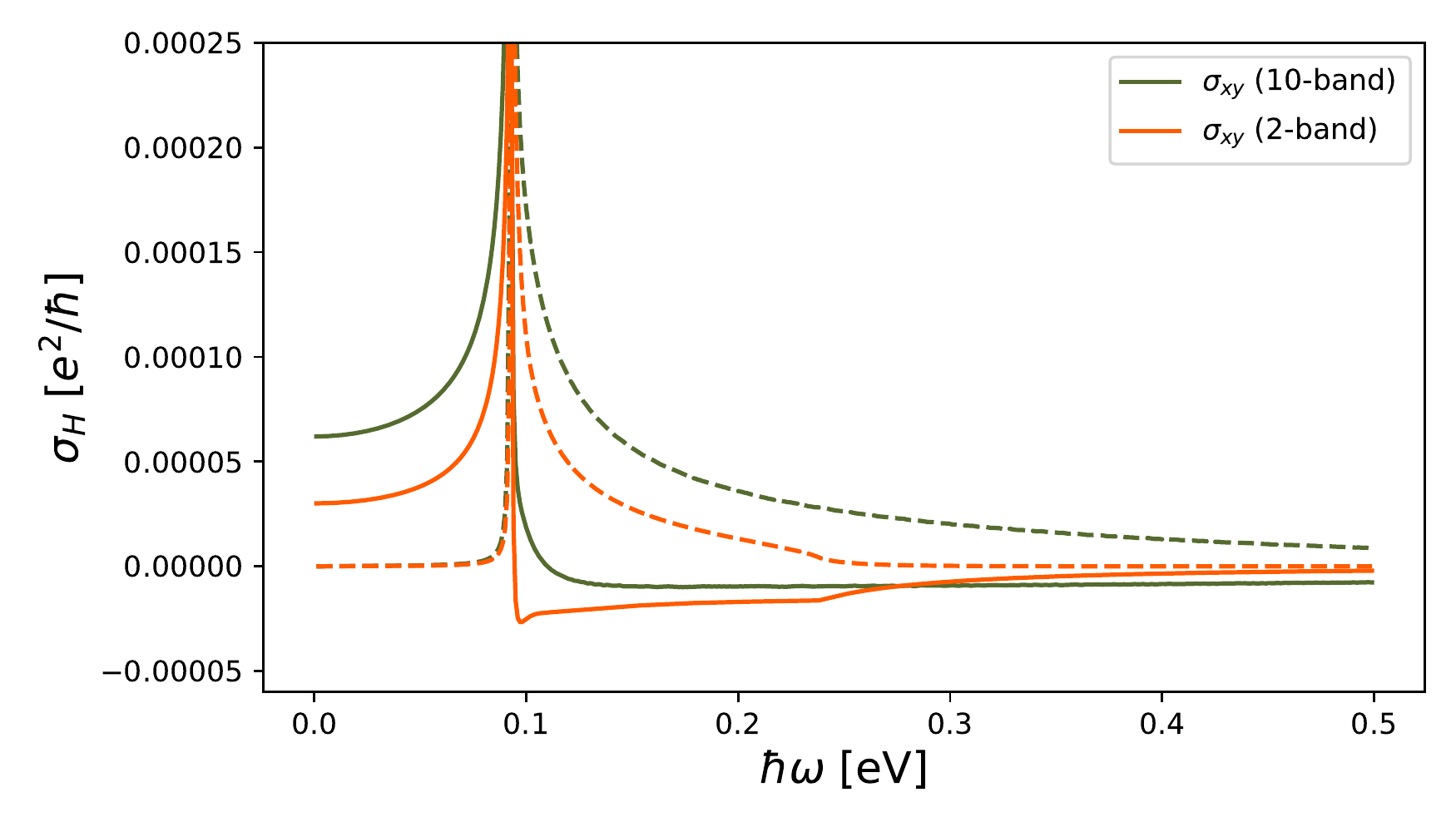}
\caption{Comparison of optical Hall conductivities $\sigma_{xy}(\omega)$ between the 2-band (shown in orange) and the 10-band (green) models. Solid and dashed lines show real and imaginary parts respectively. For the 10-band model, tight binding parameters are chosen to be $t=0.153$eV, $t'= 0$, $\mu=-3.9t$, $\Delta_{\rm max}=0.4$eV and $g_0^{(10{\rm band})}=0.008$eV. The phase between monolayers is chosen to be $\phi=0.4\pi$, breaking the time reversal symmetry. Note that in this plot we calculate the Hall conductance in a different regime and using different parameters (larger order parameter and lower chemical potential compared to the calculations in Fig.\ \ref{fig:conductance-10band-model}) to show agreement between 2-band and 10-band models.  }
\label{fig:conductance-comparison}
\end{figure}

{\em 2-band vs.\ 10-band model results comparison. --} In Fig.\ \ref{fig:conductance-comparison} we contrast our results for $\sigma_H(\omega)$ computed for the two models. We use parameters relevant to the highly overdoped (low filling) regime where we expect some degree of agreement in the optical conductivity between the two models. The tight binding parameters are related as indicated above Eq.\ \eqref{eq:bdgham_twoband}. The main difference between the two Hamiltonians is the assumed form of the interlayer couplings. In the 10-band model we assume $g_{ij}$  exponentially decaying with distance as given in Eq.\ \eqref{eq:10band_interlayercouplingform} of the main text, while in the 2-band model we use $g(\bk)=4g_0\cos{(k_xb/2)}\cos{(k_yb/2)}$ corresponding to a nearest neighbor hopping between two monolayers offset by $(b/2,b/2)$. 
We fix the amplitude of the interlayer coupling in the 10-band model and adjust $g_0$ in the 2-band model such that the onset frequencies of the ${\rm Im}\sigma_H(\omega)$ for two models match. For $g_0^{\rm (10band)}=8$meV this happens when $g_0^{\rm (2band)}\approx 1.67g_0^{\rm (10band)}$. 

Results in Fig.\ \ref{fig:conductance-comparison} show that over the entire frequency range the behavior of $\sigma_H(\omega)$ is qualitatively similar in the two models, with the amplitudes differing by about a factor of 2. This level of agreement confirms that the two-band model captures the essential physics of the bilayer system.  The discrepancy in the amplitude can be attributed to the difference in the form of the interlayer coupling and to the mismatch between the two band structures away from the BZ center. 
\subsection{Kerr angle estimation details}
\begin{figure}[t]
    \includegraphics[width=\columnwidth]{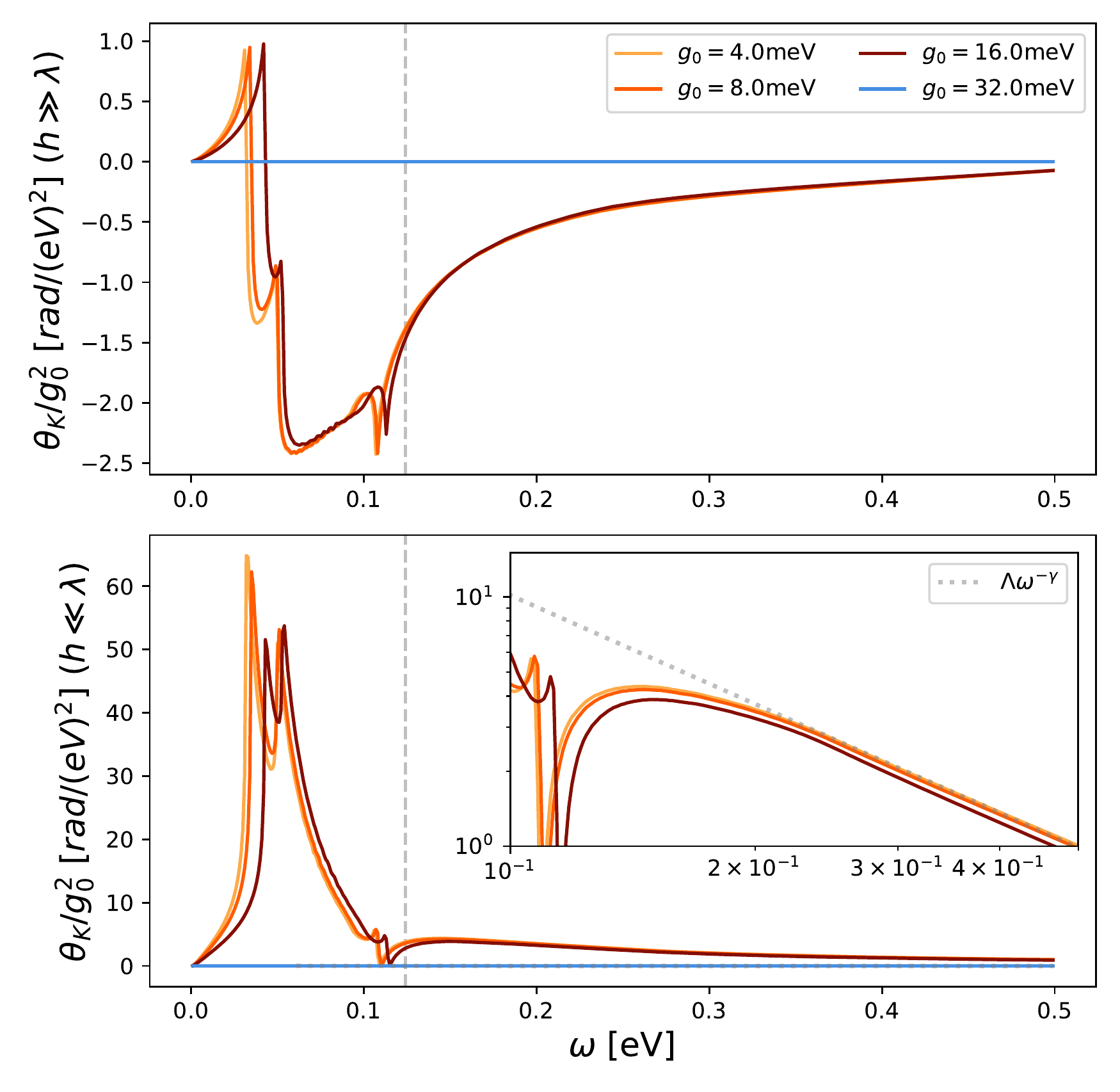}
        \caption{Scaling collapse of $\theta_K/g_0^2$ indicating leading $g_0^2$ dependence of the Kerr angle on the interlayer coupling. Top and the bottom panels show thick and thin sample limits, respectively.}
    \label{fig:kerrangle_scaling}
\end{figure}
{\em Material properties --}
We use data in the Fig.\ 6 of Ref.\ \cite{Hwang2007} to estimate the diagonal part of the optical conductivity $\sigma(\omega)$ of the bulk Bi$_2$Sr$_2$CaCu$_2$O$_{8+\delta}$ near optimal doping. As remarked by the authors the data above $\omega_c=0.12$ eV is consistent with the power law behaviour $\sigma(\omega)=C(-i\omega)^{\gamma-2}$. The measured quantity is $|\sigma(\omega)|=C\omega^{\gamma-2}$ and we extract the coefficient $C$ and the exponent $\gamma$ by considering two points on the graph $p_1=(10^3,2\times10^3)$ and $p_2=(10^4,5.6\times10^2)$ in units of $([\text{cm}^{-1}],[\Omega^{-1}\text{cm}^{-1}])$.  We find that $C=9.122\times 10^4\Omega^{-1}(\text{cm}^{-1})^{3-\gamma}$ and $\gamma=1.447$. 

For the index of refraction $n$ appearing in Eq.\ \eqref{eq:kerrangle} we use \cite{Hwang2007}
\begin{equation}
    n=\sqrt{\epsilon_H+\frac{i\sigma(\omega)}{\epsilon_0\omega}}
\end{equation}
where $\epsilon_H=4.77$  is the background dielectric tensor. We estimate the 2D conductivity $\sigma_{xx}$ (required in Eq.\ \ref{eq:thinfilm_kerrangle}) from the bulk conductivity through $\sigma = \sigma_{xx}/2d$ where $d=12.6$\AA is the interlayer spacing \cite{Can2020} between Cu-O planes. Note that we consider two Cu-O monolayers in our 2D sample.

{\em Scaling properties of the Kerr angle --} Here we discuss the dependence of the Kerr angle on the interlayer coupling strength $g_0$ and frequency $\omega$ focusing specifically on the the experimentally relevant  high frequency region. We observe a scaling collapse of the curves in Fig. \ref{fig:kerrangle_scaling}, valid at small $g_0$, when we plot $\theta_K/g_0^2$ in both thin and thick sample limits. This is consistent with the expected leading quadratic dependence on the interlayer coupling found in the 2-band model. 

\begin{figure}[b!]
    \includegraphics[width=\columnwidth]{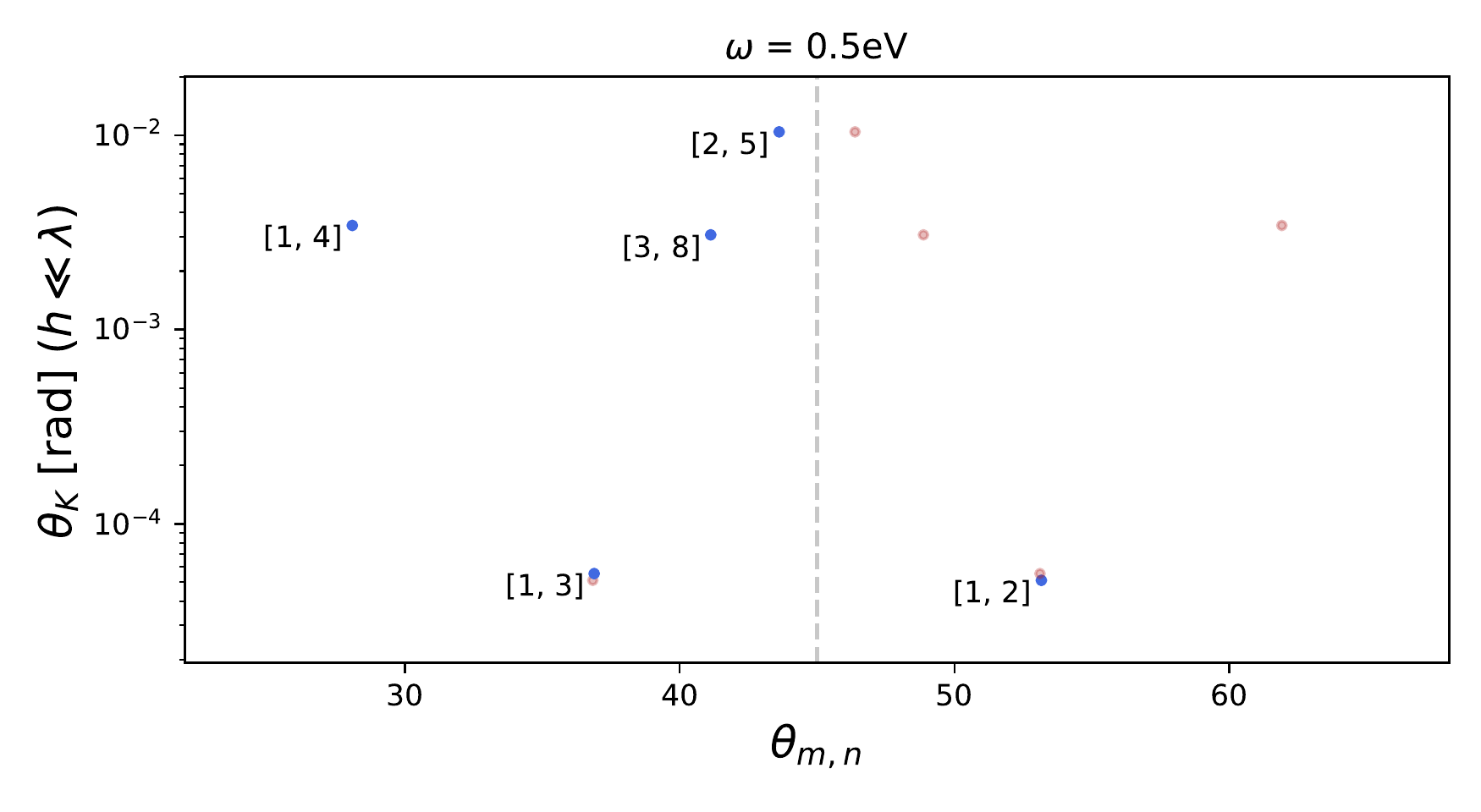}
        \caption{Kerr angle estimates for other commensurate twist angles $\theta_{m,n}$ at $\omega = 0.5$eV in the thin sample limit ($h \ll \lambda$), annotated by the twist vector $[m,n]$ (blue dots). Each data point is mirrored (red dots) with respect to $\theta_{m,n}=45^{\circ}$  to show the expected behaviour for the entire range of twist angles. }
    \label{fig:kerrangle_othercommensurate}
\end{figure}
At high frequencies comparison of Fig.\ 1 and Eq.\ \eqref{eq:timusk} indicates that  $\sigma_{xy}$ is smaller than $\sigma_{xx}$ by at least three orders of magnitude (comparing both real and imaginary parts). It is also true that, at high frequencies $|{\rm Re}\sigma_{xy}| \ll |{\rm Im}\sigma_{xy}|$ and $\text{Re}\sigma_{xy}\propto \omega^{-2}$. Therefore, in Eq.\ \eqref{eq:thinfilm_kerrangle}, it is permissible to neglect $\sigma_{xy}$ compared to $\sigma_{xx}$, which leaves us with
\begin{equation}
    \theta_K\simeq\text{Re}\arctan\left(\frac{-\sigma_{xy}}{\sigma_{xx}+4\pi\sigma_{xx}^2}\right).
\end{equation}
Because $\sigma_{xx}$ shows $\propto \omega^{\gamma-2}$ scaling with $\gamma<2$, at large enough frequencies we can drop the $\sigma_{xx}^2$ term as well. Finally, combining the frequency dependences of $\sigma_{xy}$ and $\sigma_{xx}$ and noting that $\sigma_{xy}/\sigma_{xx} \ll 1$, we expand $\arctan{x} \approx x$ for $|x| \ll 1$ and obtain Eq.\ \eqref{scaling} quoted in the main text. Therein $\Lambda$ is a prefactor dependent on microscopic details and we included the $g_0^2$ factor to capture the leading dependence on the interlayer coupling strength observed in Fig.\ \ref{fig:kerrangle_scaling}.

{\em Results for other commensurate angles ---} Our lattice model calculations so far focused on a single commensurate twist angle $\theta_{1,2}$. We have also estimated the Kerr angle in the thin sample limit for other commensurate angles which are computationally accessible and  Fig.\ \ref{fig:kerrangle_othercommensurate} shows these results.  We find that the Kerr angle is increased as the twist angle $\theta$ approaches $45^{\circ}$. Compared to the $\theta_{1,2}$ case we studied above, $\theta_{2,5} \approx 43^{\circ}$ configuration results in a $\theta_K$ at least two orders of magnitude larger. 


\begin{widetext}
\subsection{Optical Hall conductivity formulas}
\textit{General multi-band case.} -- Here we obtain an expression for Hall conductivity in terms of the eigenspectrum of the BdG Hamiltonian $h_\bk\rvert a\bk\rangle = E^a_{\bk} \rvert a\bk\rangle$. It is convenient to define the matrix element $v^x_{a\bk_1,b\bk_2}=\langle a{\bk_1}\lvert \hat{\textbf{v}}_x\rvert b{\bk_2}\rangle$ and use a collective notation $n=a{\bk_1},m=b{\bk_2}$. 
From Eq.~\eqref{eq:jxjycorrelator} in the main text, we have
\begin{equation}\label{eq:jj}
\begin{split}
    \pi_{xy}(\bq,\omega) 
    &= \int _0^{\infty} d t e^{i \omega t} \,\mathrm{tr} \left[\hat\rho \,[\hat{j}_{x}^\dag(\bq,t), \hat{j}_{y}(\bq,0)] \right]  \\
    &= \frac{1}{Z}\int _0^{\infty} d t e^{i \omega t} \, \sum_{nml} \left[\langle l\lvert e^{-\beta H} \rvert n\rangle \langle n\lvert \hat{j}_{x}^\dag(\bq,t) \rvert m\rangle \langle m\lvert \hat{j}_{y}(\bq,0) \rvert l\rangle - \langle m\lvert e^{-\beta H} \rvert l\rangle \langle l\lvert \hat{j}_{y}(\bq,0) \rvert n\rangle \langle n\lvert \hat{j}_{x}^\dag(\bq,t) \rvert m\rangle \right]  \\
    &= i \sum_{mn} \frac{e^{-\beta E_n}-e^{-\beta E_m}}{Z} \frac{\langle n \lvert \hat{j}_{x}^\dag(\bq)\rvert m\rangle  \langle m \lvert \hat{j}_{y}(\bq)\rvert n\rangle}{(\omega +i \epsilon)+E_n-E_m}\\
    &=ie^2 \sum_{\bk,ab}  v^{x\dag}_{a\bk+\bq,b\bk} v^y_{b\bk,a\bk+\bq}  \frac{n_F(E^a_{\bk})-n_F(E^b_{\bk+\bq})}{(\omega +i \epsilon) + E^a_{\bk} - E^b_{\bk+\bq}}
\end{split}
\end{equation}
In the second line, we use eigenstates $h \rvert n\rangle = E_n \rvert n\rangle$ to express the trace, the partition function $Z=\mathrm{tr}(e^{-\beta h})$, and the density matrix $\hat\rho=e^{-\beta h}/Z$. 
In the third line, we used the time dependence of the current operator $\langle n \lvert \hat{j}_{\alpha}(t) \rvert m\rangle 
= \langle n \lvert e^{i h t} \hat{j}_{\alpha} e^{-i h t} \rvert m\rangle  
=  e^{i (E_n - E_m) t} \langle n \lvert \hat{j}_{\alpha}\rvert m\rangle$ and performed the time integral.
Noting that $\hat{\textbf{j}}(\bq)=e\sum_{a\bk} (\bk+\frac{\bq}{2}) c_{a\bk+\bq}^\dag c_{a\bk}$ shifts the electron momentum by $\bq$, we immediately see that $\langle n \lvert \hat{j}_{\alpha}(\bq)\rvert m\rangle = \langle a\bk_1 \lvert \hat{j}_{\alpha}(\bq)\rvert b\bk_2 \rangle \delta_{\bk_1,\bk_2+\bq}$, which  justifies the fourth line. 
In addition, we identify the Fermi distribution $n_F(E_n)$ with $e^{-\beta E_n}/Z$ because we are dealing effectively with a noninteracting system.
For the $\bq=0$ case of interest, we have $\hat{\textbf{v}}_\alpha (\bq=0) = (\mathbbm{1} \otimes \sigma_z)\partial_{k_\alpha}h^0_\bk$ in the first quantized form of the original orbital basis, suitable for numerical evaluation.

An alternative way to derive the final expression in Eq.\ \eqref{eq:jj} is to perform an $S$-matrix expansion of the current-current correlator to one-loop level  \cite{Kallin2012} obtaining
\begin{equation}\label{pi_hallconductivity}  \pi_{xy}(\bq,\nu_m)=\frac{ie^2}{\beta}\sum_{\bk,\omega_n}\text{tr}\left[\hat{\textbf{v}}_x{(\bk+\frac{\bq}{2})} G_0(\bk,\omega_n)\hat{\textbf{v}}_y{(\bk+\frac{\bq}{2})} G_0(\bk+\bq,\omega_n+\nu_m)\right]
\end{equation}
where $G_0(\bk,\omega_n)=(-i\omega_n + h_\bk)^{-1}$  is the Green's function. Expressing $G_0$ in the spectral representation
\begin{equation}\label{spectralrepresentation}
    G_0(\bk,\omega_n)=\int d\omega  \frac{\zeta_{\bk}(\omega)}{-i\omega_n + \omega}
  \end{equation}
 with the matrix $\zeta_{\bk}(\omega)$ containing all orbital degrees of freedom
 and substituting into Eq.\ \eqref{pi_hallconductivity} we obtain
\begin{equation*}
    \pi_{xy}(\bq,\nu_m)=ie^2\int d\omega d\nu \sum_{\bk}\text{tr}\left[\hat{\textbf{v}}_x \zeta_{\bk}(\omega)\hat{\textbf{v}}_y \zeta_{\bk+\bq}(\nu)\right]\frac{1}{\beta}\sum_{\omega_n}\frac{1}{i\omega_n-\omega}\frac{1}{i\omega_n+i\nu_m-\nu}.
  \end{equation*}
As we evaluate this Matsubara sum, we note that the bosonic Matsubara frequencies $i\nu_m$ appearing in the Fermi function drop out, $n_F(\omega)=n_F(\omega-i\nu_m)$, which leads to 
\begin{equation}\label{s10}
    \pi_{xy}(\bq,\nu_m)=ie^2\int d\omega d\nu \sum_{\bk}\text{tr}\left[\hat{\textbf{v}}_x \zeta_{\bk}(\omega)\hat{\textbf{v}}_y \zeta_{\bk+\bq}(\nu)\right] \frac{n_F(\omega)-n_F(\nu)}{i\nu_m +\omega - \nu}.
\end{equation}
Since by definition $U^\dag_\bk G_0 U_\bk =\hat{\delta}(\omega-E_\bk)$ where  $[\hat{\delta}(\omega-E_\bk)]_{ij}= \delta_{ij}(-i\omega + E^i_\bk)^{-1}$, and  $U_\bk$ is a unitary matrix that diagonalizes $h_\bk$, we see that $U^\dag_\bk\zeta_\bk(\omega)U_\bk=\hat{\delta}(\omega-E_\bk)$.
We use this when inserting the identity $U^\dag_\bk U_\bk=1$ into the trace to find
\begin{equation*}
    \pi_{xy}(\bq,\nu_m)=ie^2\int d\omega d\nu \sum_{\bk}\text{tr}\left[U^\dag_{\bk+\bq}\hat{\textbf{v}}_x U_\bk \hat{\delta}(\omega-E_\bk) U^\dag_\bk\hat{\textbf{v}}_y U_{\bk+\bq} \hat{\delta}(\nu-E_{\bk+\bq}) \right] \frac{n_F(\omega)-n_F(\nu)}{i\nu_m +\omega - \nu  }.
  \end{equation*}
Taking the trace explicitly and evaluating the integrals with the help of delta functions we obtain the expression on the last line of Eq.~\eqref{eq:jj} again.

We now write down the Hall conductance \eqref{hallconductance} in the limit $q\rightarrow 0 $
\begin{align}\label{polarhallconductances1}
    \sigma_H(\omega)=\frac{ie^2}{2\omega} \sum_{\bk,ab} Q_{ab} \frac{n_F(E^a_{\bk})-n_F(E^b_{\bk })}{\omega + E^a_{\bk} - E^b_{\bk} + i\epsilon}
\end{align} 
where $Q_{ab}=\left(    v^x_{ab} v^y_{ba} - v^y_{ab} v^x_{ba} \right)_\bk$. 
We may  rewrite \eqref{polarhallconductances1} as follows
\begin{align}
    \sigma_H(\omega)=&\frac{i}{2\omega}e^2 \sum_{\bk,ab}Q_{ab}  \frac{n_F(E^a_{\bk})}{\omega + E^a_{\bk} - E^b_{\bk} + i\epsilon} - Q_{ba}  \frac{n_F(E^a_{\bk })}{\omega + E^b_{\bk} - E^a_{\bk} + i\epsilon} \\
       =&\frac{i}{2\omega}e^2 \sum_{\bk,ab}  \frac{(Q_{ab}+Q_{ba})(E^a_{\bk} - E^b_{\bk}) - (\omega + i\epsilon)(Q_{ab}-Q_{ba})}{(E^a_{\bk} - E^b_{\bk})^2 - (\omega + i\epsilon)^2 } n_F(E^a_{\bk}).
    \end{align} 
 which leads to Eq.\ \eqref{eq:sigma_H_comp} of the main text once we  account for antisymmetry $Q_{ab}=-Q_{ba}$.
Hermiticity $v^{x*}_{ab}=v^{x}_{ba}$  further implies that the antisymmetric quantity $Q_{ab} = 
2i\,\text{Im}\{v^x_{ab} v^y_{ba}\}_\bk$ is purely  imaginary.
The zero-frequency limit of the real part can thus be written as
\begin{align*}
     \lim_{\omega\rightarrow 0} \sigma_H=-ie^2 \sum_{\bk,ab}  \frac{Q_{ab} n_F(E^a_{\bk})}{(E^a_{\bk} - E^b_{\bk})^2 +  \epsilon^2 }. 
\end{align*}

\textit{Explicit formula for the 2-band model.} --
We start from the general two-band BdG Hamiltonian $h_\bk$ in the following form
\begin{equation}\label{eq:bdgham_twoband0}
    h_\bk = \begin{pmatrix} \xi_1(\bk) & \Delta_{1}(\bk) & g(\bk) & 0 \\ \Delta^*_{1}(\bk) & -\xi_1(-\bk) & 0 & -g^*(-\bk) \\
     g^*(\bk) & 0 & \xi_2(\bk) & \Delta_{2}(\bk) \\
    0 & -g(-\bk) & \Delta^*_{2}(\bk) & -\xi_2(-\bk) \end{pmatrix}.
\end{equation}
Once we assume intralayer inversion symmetry $\xi_i(-\bk)=\xi_i(\bk)$, it becomes Eq.~\eqref{eq:bdgham_twoband}. Now we have to further specify the relation between $g(\bk)$ and $g(-\bk)$.
In general, if we work with interlayer coupling that does not explicitly break $\mathcal{T}$, the complex phase of $g(\bk)$ only enters via the  $e^{i(\br_{1i}-\br_{2j})\cdot\bk}$ factors (where $\br_{aj}$ are positions of atoms in the unit cell) and thus $g^*(-\bk)=g(\bk)$. Then we have 
\begin{equation}\label{eq:bdgham_twoband1}
    h_\bk = \begin{pmatrix} \xi_1(\bk) & \Delta_{1}(\bk) & g(\bk) & 0 \\ \Delta^*_{1}(\bk) & -\xi_1(\bk) & 0 & -g(\bk) \\
     g^*(\bk) & 0 & \xi_2(\bk) & \Delta_{2}(\bk) \\
    0 & -g^*(\bk) & \Delta^*_{2}(\bk) & -\xi_2(\bk) \end{pmatrix}.
\end{equation}
For this 2-band model the trace indicated in Eq.\ \eqref{pi_hallconductivity} can be explicitly performed.
This leads to the following expression for  the Hall conductivity 
\begin{equation}
\sigma_H(\omega)= \frac{\ii e^2}{2\omega\beta} \sum_{\bk,\omega_n}
\frac{2\nu_m(\nu_m+2\omega_n)}{\left(E_-^2+\omega_n^2\right) \left(E_+^2+\omega_n ^2\right)} 
\frac{(A+B)}
{\left[E_-^2+(\nu_m +\omega_n )^2\right] \left[E_+^2+(\nu_m +\omega_n )^2\right]}\bigg\rvert_{\nu_m\rightarrow\omega+i\epsilon}
\end{equation}
where 
\begin{equation}
  \begin{split}
A&=E_+(W-E_-X)\left[|g|^2+\omega_n(\nu_m+\omega_n)-E_1E_2\right] 
-W(E_1|\Delta_2|^2+E_2|\Delta_1|^2)+\mathrm{Im}[\Delta_1^*\Delta_2]U(\nu_m+2\omega_n),\\
B&=X\left\{ \mathrm{Re}\left[ E_+E_-\Delta_1\Delta_2^*+\Delta_1^*\Delta_2^*(\Delta_1^2-\Delta_2^2) \right] \right.
\left. + (|\Delta_1|^2-|\Delta_2|^2) \left[ |g|^2-\omega_n(\nu_m+\omega_n)+E_1E_2 \right] \right\},
\end{split}
\end{equation}
with 
\[X=\left(\partial_\bk g\times\partial_\bk g^*\right)_z=2\ii\,\mathrm{Im}[\partial_{k_x}g\partial_{k_y}g^*],\ \ \ 
W=2\ii \left(\delta\bv\times\mathrm{Im}[g^*\partial_\bk g]\right)_z, \ \ 
U= \left(\delta\bv\times\partial_\bk |g|^2]\right)_z. \]
We defined the velocity anisotropy of the single-particle spectrum $\delta\bv=\partial_\bk(\xi_2-\xi_1)$ and $E_\pm=E_1\pm E_2$ with $ E_{1,2}$ the positive eigenvalues of $h_\bk$.
Here the energy denominators come from $\det G_0^{-1}(\bk,\omega_n)=\left(E_-^2+\omega_n^2\right) \left(E_+^2+\omega_n ^2\right)$.
Performing the requisite fermionic frequency summation and the analytic continuation leds to Eq.\ \eqref{eq:sigmaH_2band} shown in the main text.

\end{widetext}




\end{document}